\documentclass[12pt,a4paper]{article} 
\usepackage{graphicx}
\usepackage{lscape}
\usepackage{amsmath}
\usepackage{subfig}
\usepackage{xcolor}
\usepackage[normalem]{ulem}
\usepackage{multirow}
\usepackage{cite}

\begin{document}

\title{On the (non)genericity of the Kiselev spacetime}

\author{\.{I}brahim Semiz \\
\small Physics Department, Bo\u gazi\c ci University \\ 
\small 34342 Bebek, \.Istanbul, Turkey } 
\date{}

\maketitle

\begin{abstract}
In many works in the literature, the Kiselev spacetime is interpreted as the ``blackhole surrounded by quintessence''; an interpretation not really tenable. For example, it has recently been pointed out that the expressions ``perfect fluid'' and ``quintessence'' are used improperly in most of the articles that build on that spacetime; hence it is not as relevant as many of those authors might think. Here we point out that the original derivation of the Kiselev spacetime involves the use of too many coordinate conditions as well, again strongly restricting its relevance.
\end{abstract}

\vspace{5 mm}
In 2003, Kiselev came up with an interesting metric \cite{Kiselev}. It reads
\begin{equation}
ds^{2} = -\left[1-\frac{r_g}{r}-\sum_n \left(\frac{r_n}{r}\right)^{3w_n+1} \right] dt^{2} + \frac{dr^{2}}{\left[1-\frac{r_g}{r}-\sum_n \left(\frac{r_n}{r}\right)^{3w_n+1} \right]} + r^{2} d\Omega^{2},   \label{KiselevMetric}
\end{equation}
where we use the space-positive sign convention for the line element, unlike \cite{Kiselev}. This spacetime is sourced by a number of {\it anisotropic} fluids which are counted by the index $n$ and whose equations of state are
\begin{eqnarray}
p_{r,n} & = & -\rho_n \\
p_{t,n} & = & \frac{1+3w_n}{2}\rho_n  
\end{eqnarray}
where $\rho_n$, $p_{r,n}$ and $p_{t,n}$ are the density, radial pressure and tangential pressure, respectively, of the $n^{\rm th}$ fluid. $r_n$ are arbitrary parameters representing the contribution of the $n^{\rm th}$ fluid to the source. Thus, a metric function (and the inverse of another) contains a linear superposition of contributions of of arbitrary strength due to source fluids, a rarity indeed in the profoundly nonlinear General Relativity (GR)! This is in the present author's opinion, the most interesting aspect of the Kiselev spacetime.

However, the Kiselev spacetime is popular for another reason: Many authors seem to assume that the Kiselev spacetime, especially with one source fluid, represents a ``blackhole surrounded by quintessence''. This is an attractive concept: The discovery of the accelerated expansion of the universe tells us that it might be dominated by quintessence, hence such a blackhole would be more realistic than the asymptotically flat Schwarzschild blackhole.

On the other hand, the use of the expressions {\it perfect fluid} and {\it quintessence} in most of the many articles citing \cite{Kiselev} is incompatible with the terminology as understood in the cosmology community; as forcefully expounded in \cite{Visser_on_Kiselev} (also mentioned in \cite{Lake_on_Kiselev}): Any perfect fluid, including quintessence, must be isotropic, and the source supporting the Kiselev spacetime is not. Article \cite{Visser_on_Kiselev} points out that even though the original paper \cite{Kiselev} did state this anisotropy, the inappropriate uses of these two expressions start with that paper. 

Here we would like to point out an alternative way of seeing that the Kiselev spacetime is not as general or as relevant as most articles citing it imply; focusing on the metric {\it ansatz} used to derive it. The problem is with the ``principle of additivity and linearity'', eq.(9) of \cite{Kiselev}, equivalent to 
\begin{equation}
g_{tt} \; g_{rr} = -1,   \label{KiselevCond}
\end{equation}
which the author claims is possible to adopt without loss of generality, in the statement immediately following eq.(9). This is incorrect: 

It is well known \cite{mtw} that the general diagonal static spherically symmetric metric can be written as
\begin{equation}
ds^{2} = -B(r) dt^{2} + A(r) dr^{2} + y(r)^{2} d\Omega^{2},   \label{gen-ansatz}
\end{equation}
if one has not yet specified the choice for the radial coordinate. {\it One} such choice can be made now, e.g. specifying one of the functions in (\ref{gen-ansatz}) [{\it only one} choice, since there is only one relevant coordinate, namely, $r$]. By far the most popular choice is $y(r) = r$, called Schwarzschild coordinates or curvature coordinates. Another popular choice is isotropic coordinates, corresponding to $y(r)^{2} = A(r) r^{2}$. Some choices, including these, are shown in Table \ref{coordTable}, in terminology used by Visser and collaborators (e.g. see \cite{VisserCoordTerm}).
\begin{table}[h]
\caption{Some coordinate choices for static spherically symmetric 3+1 dimensional spacetimes. The names are those used by Visser and collaborators \cite{VisserCoordTerm}}
\begin{tabular*}{\textwidth}{p{0.18 \textwidth} p{0.18 \textwidth} p{0.55 \textwidth}}
\hline
Name & Choice & Comment/Explanation ($r$ chosen such that...)\\
\hline
Schwarzschild or Curvature & $y(r) = r$ &  the area of a sphere of symmetry is $4\pi r^{2}$\\
Isotropic & $y(r)^{2} = A(r) r^{2}$ & the spatial part of the metric is conformal to 3-D flat (Euclidean) space\\
Gaussian polar & $A(r) = 1$ & $\Delta r =$ proper distance (on a radial path)\\
Synge isothermal & $A(r) = B(r)$ & radial null trajectories satisfy $\Delta r = \pm \Delta t$\\
Buchdahl & $A(r) = B(r)^{-1}$ & correct signature is guaranteed for all $r$\\
\hline
\end{tabular*} \label{coordTable}
\end{table}

Note that a choice leaves two independent functions in the metric to be found for a solution. For example, if the spacetime is filled with a perfect fluid, this also brings two functions, namely the energy density $\rho$ and the pressure $p$. For these metric {\it ans\"atze}, the Einstein Equations give three independent components, and together with an equation-of-state of the form $f(\rho,p)=0$, we have four equations with four functions to be determined; i.e. a well-defined problem. Therefore it is entirely appropriate for the static spherically symmetric metric to contain two independent functions at the beginning of an analysis.

It is this feature that eq.(9) of \cite{Kiselev} violates: The {\it ansatz}, eq.(1) of \cite{Kiselev}, already makes the Schwarzschild choice, {\it and there is no more coordinate freedom left!} Adopting eq.(9) of \cite{Kiselev}, equivalent to (\ref{KiselevCond}) or $A(r) B(r) = 1$, on top of eq.(1) of \cite{Kiselev} decreases
the number of metric functions to one; and far from ``no loss of generality'', severely restricts the range of possible solutions. As also stated in \cite{Visser_on_Kiselev}, it leads to a very special family of metrics \cite{Jacobson}; in particular, forcing anisotropy on the stress-energy-momentum tensor. Another way to state this loss of generality is that \cite{Kiselev} is attempting to use Schwarzschild and Buchdahl coordinates simultaneously.
 
To summarize and reiterate, the Kiselev spacetime(s) are not as generic as most of the articles that cite it and build on it assume. The assumption (9) of the original article \cite{Kiselev} makes it a very particular solution. To add to the itemized list in the conclusion of \cite{Visser_on_Kiselev},
\begin{itemize}
\item Do not take the Kiselev spacetime to represent the generic, realistic blackhole in a universe dominated by quintessence; it doesn't.
\end{itemize}

\end{document}